\documentstyle[prl,aps,twocolumn]{revtex}                    
\begin{document}                  
\draft
\title{Schwarzschild horizon and the gravitational redshift formula  }
\author{ Edward Malec $^1$ }
\address{  Institute of Physics, 
Jagiellonian University,
30-064 Krak\'ow, Reymonta 4, Poland   }

\maketitle

\begin{abstract}
The gravitational redshift formula is usually derived in the 
geometric optics approximation. In this note we consider an exact
formulation of the problem in the Schwarzschild space-time, with
the intention to clarify under what conditions this redshift law 
is valid. It is shown that in the case of shocks the radial 
component of the Poynting  vector can scale according to the  
redshift  formula, under a suitable condition. If that condition
is not satisfied, then  the effect of the backscattering can lead
to significant modifications. The obtained results imply that the
energy flux of the short wavelength radiation obeys the standard 
gravitational redshift formula while the energy flux of long waves
can scale differently, with redshifts being dependent on the
frequency.

\end{abstract}

\date{ }

\section{ Introduction}

The  gravitational redshift formula in the Schwarzschild spacetime
\cite{Landau}
\begin{equation}
\omega '=\sqrt{1-{2m\over a}\over 1-{2m\over R}}\omega 
\label{0}
\end{equation}
relates the frequency $\omega '$ detected by an   observer located at $R$
with the initial frequency
$\omega $ of a photon  emitted from an areal  distance $a$ from the
gravitational center.

Formula (\ref{1}) is derived in the approximation of geometric optics (see  
 \cite{Sachs}, \cite{Ehlers}, \cite{Schneider} ).
The aim of this paper is to present  an exact  treatment of the problem
within 
the framework of the classical wave theory; the only simplification
consists in 
neglecting the backreaction effect.
 It is assumed  that  an isolated pulse  of an  electromagnetic  wave is
emitted outward.
Its initial support is contained in the annulus $(a,b)$. Initial pulses
that
are characterized by $(b-a)/a<<1$ will be - for the sake of brevity -
referred to as  shocks. 
One can define a radial component of the
 energy flux $\hat P_R$   by a suitable spherical  projection of 
the Poynting vector.  It will be shown that if the relative width
$(b-a)/a$ 
is much smaller than   some power of the relative distance 
  $(a-2m)/a$ from the horizon then the radial  
   energy flux of a shock conforms to a relation like in Eq.  (\ref{0}). 
  
The order of this paper is following. Sec. II brings the main result and
outlines its proof.  Sec. III is devoted to the estimation of a reduced
electromagnetic
 potential. In Secs IV and V we derive bounds on the radiation
amplitudes. Sec. VI 
proves the desired  flux relation.  Last section is dedicated to a short
summary and discussion 
on the limitations of the redshift formula.    

\section{Main results}

The space-time geometry  is defined  by
the Schwarzschild  line element,
\begin{equation} ds^2 = - (1-{2m\over R})dt^2 +
{1\over 1-{2m\over R}} dR^2 +
R^2 d\Omega^2~,
\label{1}
\end{equation}
where $t$ is a time coordinate, $R$ is a radial
coordinate that coincides with the areal radius
and $d\Omega^2 = d\theta^2 + \sin^2\theta d\phi^2$
is the line element on the unit sphere, $0\le \phi < 2\pi $
and $0\le \theta \le \pi $.
Throughout this paper $G$, the Newtonian gravitational constant,
and $c$, the velocity of light are put equal to 1.
We define the Regge-Wheeler coordinate $r^*=R+2m\ln ({R\over 2m}-1)$
and, for the sake of concise notation, $\eta_R \equiv 1-{2m\over R}$.
  
The Maxwell equations read 
\begin{equation}
\nabla_{\mu }F^{\mu }_{\nu }=0,
\label{2}
\end{equation}
where $F_{\mu \nu }=\partial_{\mu }A_{\nu }- \partial_{\nu }A_{\mu }$ and
$A_{\mu }$ is the electromagnetic  potential. 
It is convenient to assume $A_0=0$ and the Coulomb gauge condition
 $\nabla_iA^i=0$. It is useful to follow Wheeler \cite{Wheeler} and employ 
the multipole expansion.   We will consider only the dipole
term  and, more specifically, 
 choose  the potential one-form $A=\sqrt{3/2}\sin^2\theta \Psi
(r^*,t)d\phi $.
A similar analysis with the same conlusions can be done in any multipole
order.

We seek, following    \cite{malec2000},
a  reduced dipole potential  $\Psi (r^*,t)$   in the form
$\Psi =\tilde \Psi +\delta $,
where $\delta $   satisfies the dipole  equation
\begin{equation}
(-\partial_0^2 + \partial_{r^*}^2)\delta = \eta_R  
\Biggl[ { 2\over R^2}
\delta + {6mf\over R^4}  \Biggr] .
\label{2.2}
\end{equation}
Here $\tilde \Psi (r^*,t)= \partial_0f(r^*-t) +{f(r^*-t)\over R}$,
$\partial_0\equiv \partial_t$ 
and $f$ is an arbitrary function with support in $(a, b )$.
It is   well known   that
  $\tilde \Psi $  solves Maxwell equations  in Minkowski spacetime
\cite{MTW}.  
 $f$  can be uniquely determined from initial data corresponding
 to an initially outgoing radiation.  Initially $\delta =\partial_0\delta
=0$.

The stress-energy tensor of the electromagnetic field
is   $T_{\mu }^{\nu }=(1/2)(F_{\mu \gamma }F^{\nu \gamma }-(1/4)g_{\mu
}^{\nu }
F_{\gamma \delta }F^{\gamma \delta })$ and the time-like translational
Killing vector is denoted as $\zeta $.  
 
We define the projected energy density
\begin{eqnarray}
&&\hat \rho \equiv \int_{S(R))}dS(R)T^{\mu }_0\zeta_{\mu }\sqrt{g_{RR}}=
\nonumber \\
&& 2\pi  \Biggl(  {(\partial_0\Psi )^2\over \eta_r} + \eta_r
(\partial_r\Psi )^2+{2(\Psi )^2\over r^2}\Biggr) 
\label{2.3a}
\end{eqnarray}
The  energy  $E_R(t)$ of the electromagnetic field  $\Psi $
contained in the exterior of a sphere $S(R)$ of the radius $R$ reads 
$E_R(t) =\int_R^{\infty }dr \hat \rho $. Let $n$ be the unit normal to
$S(R)$.  
  One finds  that $\partial_tE_R(t)=   n^R\hat P_R$; thus  the energy flux 
${dE\over ds}|_{R=const}$ (here $ds=\sqrt{\eta_R}dt$ is  the   proper time
interval)  
through $S(R)$  equals to $\hat P_R$.   $\hat P_R$ is the surface integral
of the normal component of the Poynting vector,
\begin{eqnarray}
 \hat P_R(R,t)\equiv \int_{S(R)}dS(R)T^{r\mu }\zeta_{\mu }= 
 -{4\pi \over   \sqrt{\eta_R}} \partial_0\Psi \partial_{r^*}\Psi .  
\label{2.5}
\end{eqnarray}
It is convenient to introduce the reduced strength field amplitudes $h_+$ 
and $h_-$  
\begin{eqnarray}
&&    h_+\equiv {1\over \eta_R}\Bigl( -\partial_0\Psi  +\partial_{r^*}\Psi
\Bigr) ,
\nonumber \\ 
&&  h_-\equiv {1\over \eta_R}\Bigl( \partial_0\Psi +\partial_{r^*}\Psi
\Bigr) .
\label{2.6}
\end{eqnarray}
$h_+$ and $h_-$ represent the outgoing radiation   and  ingoing radiation,
respectively.
Eq. (\ref{2.5}) reads, in terms of the amplitudes $h_+$ and $h_-$,
\begin{eqnarray}
&&\hat P_R(R,t)  ={\pi \over   \sqrt{\eta_R}}\Bigl(
(\eta_Rh_+)^2-(\eta_Rh_-)^2\Bigr)      .  
\label{2.5a}
\end{eqnarray}
We define 
\begin{equation}
\epsilon \equiv (b-a)/(a\eta^5_a),
\label{epsilon}
\end{equation}
a quotient  of the relative width of the pulse
by  5-th power of the relative distance $\eta $ from the event horizon.  
This quantity, with this particular power of $\eta_a$,
 is needed  in order to prove  a sufficient
condition for the validity of the redshift formula. 
It is remarkable that $\epsilon $ depends both on
  the relative width  and
 on the relative distance from the Schwarzschild horizon.  
We   comment on this point in the last section.

 One
can prove a number of properties that are valid
for shocks, provided that they satisfy the condition $\epsilon <<1$. 
  These are

i)  the amplitude     $h_+ (R)$  is well   approximated by $\tilde h_+$,
where 
\begin{equation}
\tilde h_{+}(R,t) ={1\over \eta_R}(-\partial_0+\partial_{r^*})
\Bigl( \delta +\partial_0f\Bigr) ;
\label{tildehplus}
\end{equation}

ii) the function  $\tilde h_{+}(R(t),t)$ conforms to the scaling law
$\eta_R\tilde h_{+}(R(t),t) =
\eta_a\tilde h_{+}(R(0),t=0)$;
    
iii)    $ h_- $   is negligible.
 
Let us define $\tilde \Gamma_{R_0}$ as a null geodesic directed outward 
from the point $R_0$ of the initial hypersurface and let 
  $\tilde \Gamma_{ R_0,  (R,t) }$  be a segment of $\tilde \Gamma_{R_0}$
that connects $R_0$ and $(R,t)$.
Comparing the   energy fluxes  through the spheres $S(R)$ (where $R>>2m$) 
 and the initial   $S(R_0)$ one obtains, from (\ref{2.5a}) and properties
i) - iii), that
\begin{eqnarray}
  \hat P_R(R) \approx 
    \sqrt{\eta_{R_0}\over \eta_R}    \hat P_R(R_0) ,
\label{2.9}
\end{eqnarray}
where it is assumed that $R$ and $R_0$ are connected by 
the null geodesic segment $ \tilde \Gamma_{R_0, (R,t)}$.

\section{Estimating the reduced electromagnetic potential  }

In this section we sketch briefly derivation of bounds
on the   electromagnetic potential.  There are two   
 bounds. One of them (Eqs. (\ref{3.1} and (\ref{3.2}))
applies to initial data (or a fixed Cauchy
hypersurface), and it is essentially a Sobolev-type inequality.
 The second bound (Eq. (\ref{3.11})) is proven by
the   application of  the energy method  succeeded by 
a Sobolev-type argument. A detailed calculation can be found in     
\cite{Karkowski}. 

 One  can show,
    in the case of initial data of compact support, that
      on the initial hypersurface
\begin{eqnarray}
{4\pi  f^2(R)\over r^2}\le E_a a \eta_R^2F(\tilde m,y),
\label{3.1}
\end{eqnarray}
where $A\le R\le b$ and 
\begin{eqnarray}
F(\tilde m,y) &\equiv &
 y-1+{16\tilde m^4\over 3 (-y+2\tilde m)^3}-
 {16\tilde m^4\over 3 (-1+2\tilde m)^3} +
\nonumber \\
&&
{16\tilde m^3\over (-1+2\tilde m)^2}
-{16\tilde m^3\over (-y+2\tilde m)^2}+
\nonumber\\
&& {24\tilde m^2\over -y+2\tilde m}  -{24\tilde m^2\over -1+2\tilde m }
+8\tilde m \ln {y-2\tilde m\over 1-2\tilde m}.
\label{3.2}
\end{eqnarray}
Here  $\tilde m\equiv m/a$ and $y=b/a$.  Eq. (\ref{3.1}) is a special
case of a result proven in  \cite{Karkowski}.

It is useful to proceed as in \cite{malec2000}
and define an energy  $H(R,t)$ of  the field $\delta $,
\begin{equation}
H(R,t) = \int_{R}^{\infty }dr
\Bigl(  {(\partial_0\delta )^2\over \eta_r} +  \eta_r
(\partial_r\delta)^2+2{\delta ^2 \over r^2}\Bigr) ,
\label{3.8}
\end{equation}
The upper integration bound is in fact finite, since the original support
of $f$ was 
finite and therefore  $\delta $ must also have a finite support if
$t<\infty $. 
 This is, however, irrelevant for us.  
It can be shown  \cite{Karkowski} that
\begin{eqnarray}
\sqrt{ H(a_t, t)}\le 6m \int_0^tdt
\Biggl( \int_{a_t}^{\infty }dr { f^2\eta_r
\over r^8}\Biggr)^{1/2};
\label{3.9}
\end{eqnarray}
here the external integration countour  (the  $dt$ - integral) coincides
with 
null geodesic $\tilde \Gamma_a$.
 The internal ($dr$-) integration is done on a fixed ($t=const$) 
 Cauchy hypersurface.
It is useful to replace the integration parametr $t$ by the areal
 radius $a_t$ and to
introduce  the dimesionless   variables $x=a_t/a, y=r/a$. Provided 
that is done, the 
insertion of (\ref{3.1}) into (\ref{3.9}) leads to 
\begin{eqnarray}
&&\sqrt{ H(a_t, t)}\le
\nonumber \\
&& 6{m\over a} \sqrt{E_aF(\tilde m, y)\over 4\pi }
\int_1^x{dx\over 1-{2\tilde m\over x}}
\Biggl( \int_{x}^{\infty }dy {  (1-{2m\over y})^3
\over y^6}\Biggr)^{1/2}.
\label{3.9a}
\end{eqnarray}
The internal integration can be done explicitly, so that
\begin{eqnarray}
&&\sqrt{ H(a_t, t)}\le 6{m\over a} \sqrt{E_aF(\tilde m, y)\over 4\pi }
\nonumber \\
&& \times \int_1^x{dx\over 1-{2\tilde m\over x}}
\Biggl( {1\over 5x^5}-{\tilde m\over x^6}+{12\tilde m^2\over 7x^7}-{\tilde
m^3
\over x^8} \Biggr)^{1/2}.
\label{3.9b}
\end{eqnarray}
The right hand side of (\ref{3.9b}) is bounded from above by
\begin{eqnarray}
&& 6{m\over a} \sqrt{E_aF(\tilde m, y)\over 4\pi }
\Biggl( \int_1^{\infty }{dx\over (x-2\tilde m)^2}\Biggr)^{1/2}
\nonumber \\
&&\times  \Biggl( \int_1^{\infty }dx
\Bigl[ {1\over 5x^3}-{\tilde m\over x^4}+{12\tilde m^2\over 7x^5}-{\tilde
m^3
\over x^6}\Bigr] \Biggr)^{1/2} ;
\label{3.9c}
\end{eqnarray}
here the integration has been extended to $\infty $ and the  Schwarz 
inequality  has been used. This now can be calculated exactly, but the
numerical
factors do not really matter - 
from our point of view  the only interesting fact is that 
\begin{eqnarray}
\sqrt{ H(a_t, t)}\le  C \sqrt{ F(\tilde m, y) \over \eta_a }, 
\label{3.9d}
\end{eqnarray}
where $C$ is some (time-independent) constant.
 Later on  we  will always use the same symbol $C$ in order to denote
various   numerical factors that appear in the course of calculations.
 
One can   derive a  bound
\begin{eqnarray}
{|\delta |\over  R}\le \sqrt{ 2H(a_t, t)\over R\eta_R};
\label{3.10}
\end{eqnarray}
in order to do this, observe that ${|\delta |\over  R}=|\int_R^{\infty }
dr \partial_r (\delta /r)$. The use of the Schwarz inequality yields now
immediately, taking into account  (\ref{3.8}),
${|\delta |\over  R}\le \sqrt{\int_R^{\infty }dr/(r-2m)^2}\sqrt{2H}=
\sqrt{ 2H(a_t, t)\over R\eta_R}$. 

Formulae   (\ref{3.9d}) and (\ref{3.10}) give an estimate

\begin{equation}
 {|\delta |\over \sqrt{R}} \le {C\over \eta_a} \sqrt{ F(\tilde m, y)  } .
\label{3.11}
\end{equation}

\section{Bounds on the ingoing radiation}

The amplitude $h_-$ of the ingoing radiation can be split as follows

\begin{eqnarray}
h_-\equiv   \tilde h_-  + \hat h, 
\label{3.4}
\end{eqnarray}
where 
\begin{equation}
\tilde h_- 
\equiv {1\over \eta_R}(\partial_0+\partial_{r^*})\tilde \Psi 
\label{tildehminus}
\end{equation}
and  
\begin{equation}
\hat h\equiv {1\over \eta_R}(\partial_0+\partial_{r^*})\delta .
\label{hhat}
\end{equation}
   $\tilde h_-$  represents the amplitude  related to   $\tilde \Psi $ 
   while  $\hat h$ is the term induced by the backscatter. We shall bound
$h_-$ 
 within the main flow inside that region of the Schwarzschild
 spacetime that is  bounded by null cones spanned by spherical bundles of
geodesics 
$\tilde \Gamma_a$ and $\tilde \Gamma_b$, $t>0$.

A straightforward algebra yields
 $\tilde h_- \equiv  
{1\over \eta_R}(\partial_0+\partial_{r^*})\tilde \Psi   =  
  -{f \over R^2}$. The use of (\ref{3.1}) gives  
\begin{eqnarray}
| \tilde h_- | \le   {\eta_R\over R}
\Biggl( {E_a a\over 4\pi  }  F(\tilde m, y)\Biggr)^{1/2}
\le {C\over R} \sqrt{F(\tilde m, y)}.
\label{3.6}
\end{eqnarray}

The function $\hat h$ vanishes identically at $t=0$,
by the definition of $\tilde \Psi $. Below we derive
a bound  on the function  $\hat h$.

The evolution equation (\ref{2.2}) can be rewritten   as
\begin{equation}
(-\partial_0 + \partial_{r^*})\Bigl( \eta_R\hat h\Bigr)
= \eta_R \Biggl[ {2\over R^2}
\delta + {6mf\over R^4 }
\Bigr) \Biggr] .
\label{3.12}
\end{equation}
Define a null geodesic $\Gamma_c$ that is directed inward
 from a point  $c $ of the initial hypersurface.
Notice that $\hat h$ vanishes identically along the outgoing
null  geodesic  $\tilde \Gamma_b$. 
 Taking this into account one obtains, integrating (\ref{3.12}) along
$\Gamma_c$, $c>b$, 
\begin{equation}
  \eta_R\hat h (R,t)
= \int_c^Rdr
\Biggl[ {2\over r^2}
\delta + {6mf\over r^4 }
\Bigr) \Biggr] .
\label{3.12a}
\end{equation}
The insertion of the estimate (\ref{3.11}) on the function $\delta $ 
as well of the bound  (\ref{3.1}) on $f$, allows one to deduce from
(\ref{3.12a})

\begin{equation}
 |\hat h |\eta_R \le C\sqrt{F(\tilde m, y)   \over  R\eta_a}.
\label{3.13}
\end{equation}

 \section{Scaling of the amplitude of the outgoing radiation}

First we show that in the case of shocks
$h_+$ can be approximated by $\tilde h_+$.
Indeed, from the definitions of $h_+$  
and $\tilde h_+$ ((\ref{2.6}) and
(\ref{tildehplus}), respectively) follows
\begin{equation}
h_+=\tilde h_+-{f\over R^2}.
\label{3.14}
\end{equation}
  We can conclude, taking into account the bound (\ref{3.1}), that
\begin{equation}
|h_+-\tilde h_+|={|f|\over R^2}|\propto {1\over R}\sqrt{F(\tilde m, y) }.
\label{3.14a}
\end{equation}
One can easily see, by inspection of formulae (\ref{epsilon}) and
(\ref{3.1}),
 that ${\sqrt{F(\tilde m, y) } \over \eta_a} <C\sqrt{\epsilon }$.
Therefore the   difference in (\ref{3.14a})
 becomes infinitesimally small if   $\epsilon <<1$.

The evolution equation (\ref{2.2}) can be written  in terms of $\tilde
h_+$
 as
\begin{equation}
(\partial_0 + \partial_{r^*})\Bigl( \eta_R\tilde h_+\Bigr)
= \eta_R \Biggl[ {2\over R^2}
\delta + {6mf\over R^4 }
\Bigr) \Biggr] .
\label{3.14b}
\end{equation}
Let us recall, that the right hand side of (\ref{3.14}) can be
pointwise bounded, using (\ref{3.1}) and (\ref{3.11}).
  The integration of  (\ref{3.14}) along a null
geodesic $\tilde \Gamma_{R_0}$, $a\le R_0\le b$  yields now
\begin{equation}
\eta_R\tilde h_{+}(R(t),t) -
\eta_{R_0}\tilde h_{+}(R_0)\le   C {\sqrt{F(\tilde m, y) } \over \eta_a} .
\label{3.15}
\end{equation}
Fixing the  energy  $E_a $, one notices that  by choosing 
   $\epsilon <<1$ the right hand side of (\ref{3.15}) can be made
arbitrarily small (see the remark following Eq. (\ref{3.14a})). 
Thus the product $\eta_R\tilde h_{+}$ is   constant. In  this case
one clearly sees the manifestation of the redshift - the rescaling
of the amplitude $\tilde h_{+}$.  If $R>>2m$ (a distant  observer),
then  $\tilde h_+( R )=\eta_{R_0}\tilde h_+(R, t=0)$.
 A similar  fact was discussed earlier 
 in the context of   a massless scalar field theory \cite{MENOM}.
 
\section{Calculating the flux.}

Recall the formula (\ref{2.5a}):
\begin{eqnarray}
 \hat P_R(R,t)  ={\pi  \over  \sqrt{\eta_R}}\Bigl(
(\eta_Rh_+)^2-(\eta_Rh_-)^2\Bigr)      .  
\label{4.1}
\end{eqnarray}
The estimates of the preceding section allow us to write, in the case of
shocks,
the initial value of the   flux
\begin{eqnarray}
 \hat P_R(R_0,0)  = {\pi  \over  \sqrt{\eta_{R_0}}}\Bigl(
(\eta_{R_0}\tilde h_+)^2 +  
 O(\sqrt{F(\tilde m, y)})  
\label{4.2}
\end{eqnarray}
and  (if $(r,t)\in \tilde   \Gamma_{R_0}$)
\begin{eqnarray}
 \hat P_R(R,t)  ={\pi  \over \sqrt{\eta_{R}} }\Bigl( (\eta_R\tilde h_+)^2 
 +  O(\sqrt{F(\tilde m, y)}) .
\label{4.3}
\end{eqnarray} 
One can  show, as pointed in Sec. V,  that $F(\tilde m, y) <C
\eta^2_a\epsilon $.
The product $\eta_R\tilde h_+$ is   constant if $\epsilon <<1$, as shown
in Sec V.  
Therefore one obtains, comparing (\ref{4.1}) and (\ref{4.3}), the required
relation
\begin{equation}
\hat P_R( R,t )  =\sqrt{\eta_{R_0}\over \eta_R}\hat P_R(R_0,0).
\label{4.4}
\end{equation}
 \section{Conclusions}

 We have  shown that if a relative width $(b-a)/a$ 
 of a shock is much smaller than some
 power of its relative distance $\eta_a$ from the 
 horizon of the Schwarzschild black hole,
then the   radial energy flux  
satisfies    Eq. (\ref{4.4}).
The condition that $\epsilon <<1  $   can be 
interpreted (appealing to the so-called 
similarity theorem of the Fourier transform 
theory \cite{Bracewell}) as the  demand that the 
radiation is dominated by sufficiently high
frequencies.  
Thus one infers that in the high frequency limit the backscatter 
is negligible; invoking  now to the geometric  optics  approximation,
one can relate the energy fluxes with frequencies \cite{Sachs} and  
the standard redshift formula (\ref{0}) follows.
It is necessary to point out, that barring backscatter, 
all of the energy of an outgoing pulse would get to infinity, 
although  its energy flux would have to obey (\ref{4.4}). 
The only mechanism  that  can diminish the energy
is the backscatter of the radiation  off the curvature of 
the spacetime.
 
The compactness condition $\epsilon <<1$ is sufficient but not
necessary for the validity of the standard redshift formula.
The necessary condition is probably weaker - $ (b-a)/(a\eta^k)<<1$,
where $k$ is some number not bigger  than 5 and strictly bigger than 1.   
In the case of  frequencies much lower than 
 a critical value $\eta^k_a/(b-a)$     the asymptotic energy
flux might well disobey   the  
 scaling   law (\ref{4.4}).     
 That suggests that the effect of the backscattering  can 
 be discovered   by the observation of discrete spectra.    
Any attempt to fit observed data to the simple scaling law of
 (\ref{0}) would  lead 
to  redshifts  depending on the frequency, in the case of 
a significant  backscatter.  

In the present paper the consideration is focused  only on the dipole
term, but a similar analysis with the same conlusions can be done in any 
multipole order. The key part that would require a  
modification is the Sobolev type  estimate of section III; the adaptation
of the
 remaining parts of the procedure is straightforward.

{\bf Acknowledgements.}     
  The author  thanks Bernd Schmidt and Jiri Bicak
for discussions and  all colleagues from 
the  Albert Einstein Institute of the Max Plank Institute  at Golm for
their  hospitality. A   support of the KBN grant 2 PO3B 010 16 
is acknowledged.


\begin{references}
\bibitem{*}  This work has been done during the stay of
the author at  AEI MPG at Golm.
\end{references}
\end{document}